# Excitations in the spin-1 trimer chain compound $CaNi_3P_4O_{14}$: gapped dispersive spinwaves to gapless magnetic excitations

A. K. Bera[1] and S. M. Yusuf[1,2,*]

[1]Solid State Physics Division, Bhabha Atomic Research Centre, Mumbai 400085, India
[2]Homi Bhabha National Institute, Anushaktinagar, Mumbai 400094, India

D. T. Adroja[3,4]

[3]ISIS Facility, STFC Rutherford Appleton Laboratory, Harwell Oxford, Didcot OX11 0QX, United Kingdom
[4]Highly Correlated Matter Research Group, Department of Physics, University of Johannesburg, Auckland Park 2006, South Africa

**Abstract**

Magnetic excitations and the spin Hamiltonian of the spin-1 trimer chain compound $CaNi_3P_4O_{14}$ have been investigated by inelastic neutron scattering. The trimer spin chains in $CaNi_3P_4O_{14}$ are resulted from the crystal structure that provides a special periodicity of the exchange interactions ($J_1$-$J_1$-$J_2$) comprised of intratrimer ($J_1$) and intertrimer ($J_2$) exchange interactions along the crystallographic $b$ axis. Experimental data reveal gapped dispersive spin wave excitations in the 3D long-range ordered magnetic state ($T_C$ = 16 K), and gapless magnetic excitations above the $T_C$ due to the low dimensional spin-spin correlations within chains. Simulated magnetic excitations, by using the linear spin wave theory, for a model of coupled trimer spin-chains provide a good description of the observed experimental data. The analysis reveals both ferromagnetic $J_1$ and $J_2$ interactions within the chains, and an antiferromagnetic interchain interaction $J_3$ between chains. The strengths of the $J_1$ and $J_2$ are found to be closer ($J_2/J_1 \sim 0.81$), and $J_3$ is determined to be weaker ($J_3/J_1 \sim 0.69$), which is consistent with the spin chain type crystal structure. Presence of a weak single-ion-anisotropy ($D/J$=0.19) is also revealed. The strengths and signs of exchange interactions explain why the 1/3 magnetization is absent in the studied spin-1 compound $CaNi_3P_4O_{14}$ in contrast to its $S$= 5/2 counterpart Mn based isostructural compound. The signs of the exchange interactions are in agreement with that obtained from the reported DFT calculations, whereas, their strengths are found to be significantly different. The relatively strong value of the $J_3$ in $CaNi_3P_4O_{14}$ gives a conventional 3D type magnetic ordering behaviour below the $T_C$ with full ordered moment at 1.5 K, however, retains its 1D character above the $T_C$. The present experimental study also reveals a sharp change of single-ion anisotropy across the $T_C$ indicating that the stability of the 3D magnetic ordering in $CaNi_3P_4O_{14}$ is ascribed to the local magnetic anisotropy in addition to interchain interactions. The present study divulges the importance of full knowledge of the exchange interactions in trimer spin chain compounds to understand their exotic magnetic properties, such as 1/3 magnetization plateau. The importance of the observed gapless magnetic excitations of the strongly correlated spin-chains above the $T_C$ is also discussed.

* Corresponding author, Email: smyusuf@barc.gov.in

**INTRODUCTION:**

Low dimensional magnetic materials, in particular one dimensional (1D) spin-chains, are currently of great interest in condensed matter physics as model experimental systems to study the physics of many-body quantum physics [1]. In general, the magnetic ordering in 1D spin systems is suppressed even at $T = 0$ K by strong quantum fluctuations. Further the ground-state and low-lying excitations of half-integer spin-chains are completely different from those for integer spin-chains [2, 3]. Coupled spin-chains are of further interest as the interchain couplings lead to unusual magnetic properties, and can even stabilize magnetic ordering (beyond the critical value of the interchain coupling) with diverse characteristics [4, 5]. Further, spin-chain systems consist of multiple intrachain interactions, such as alternating [6], dimer [7], trimer [8], and tetramer [9] spin-chains have recently attracted much attention due to their various unconventional magnetic properties that originate from the periodicity of exchange interactions within the chains [10].

Among the aforementioned systems, the trimer spin-chains are of special interest to us. Trimer units are formed by three consecutive atomic spins which are coupled by strong intratrimer exchange interactions $J_1$'s. A trimer spin-chain is formed by interconnecting trimers in the one dimension with intertrimer exchange interaction $J_2$ which leads to periodic exchange interactions $J_1$-$J_1$-$J_2$. Among the diverse properties of trimer spin chains, the occurrence of the magnetization plateaus, that can be viewed as an essentially macroscopic quantum phenomenon, has gained much attention recently. Analogous to the quantum Hall effect, at the magnetization plateau state magnetization is quantized to fractional values of the saturated magnetization value, proving a striking example of the macroscopic quantum phenomenon. In the trimer spin chain systems, the presence of magnetization plateaus is theoretically predicted for a wide range of interactions strengths of $J_1$ and $J_2$ as well as spin values where the number of the plateaus ($2S+1$) strongly depends on the spin value $S$ [11]. The appearance of the plateaus is theoretically predicted for both ferromagnetic (FM) and antiferromagnetic (AFM) intratrimer interactions $J_1$ with FM-FM-AFM and AFM-AFM-FM/AFM configurations [11]. It is also predicted that apart from the signs of the $J_1$ and $J_2$, nature of magnetic ground state and the appearance of magnetization plateau states are also dependent on the relative strength ($J_2/J_1$) of the intrachain interactions, interchain interactions ($J_3$), spin values as well as anisotropy in trimer spin-chains. The ground state for an AFM trimer chain (AFM $J_1$) with an integer spin is theoretically predicted to be trimerized and a gap is predicted between the ground state and the first excited state for an extended range of exchange interactions strengths [12]. On the other hand, the ground state for the half-integer counterpart is degenerate [13]. It was theoretically reported that, for AFM-AFM-FM trimer chain the magnetization plateaus appear only below a critical value of ($J_2/J_1$), whereas, for FM-FM-AFM trimer spin chain, plateaus appear above a critical value of ($J_2/J_1$). Both the cases, the essential condition for the plateau state to be observed is the dominating AFM interactions. Therefore, knowledge of the sign and strength of the exchange interactions $J_1$ and $J_2$ is essential for a trimer spin-chain compound to understand microscopically the presence or absence of magnetization plateaus as well as their nature. It is also essential to know the ratios of the exchange interactions ($J_1$, $J_2$ and $J_3$) which determine how good or bad a system to be considered as a trimer system as well as to correlate with their physical properties.

However, experimental studies on model spin-trimer chain compounds are limited. Among the experimentally reported compounds, the spin-1/2 AFM trimer chain compounds $A_3Cu_3(PO_4)_4$ ($A$=Ca, Sr, and Pb) exhibit 1/3 magnetization plateaus [14]. The nature and strengths of the exchange interactions were estimated from the temperature dependent susceptibility, and the compounds were defined to be good trimer systems with $J_2/J_1$ = 0.02-0.11. The magnetic ordering temperatures for these compounds were found to be low; $T_c$ = 0.91 K for Ca, $T_N$ = 0.91 K for Sr and $T_N$ = 1.26 K for Pb compounds, respectively. The 1/3 magnetization plateau was also reported for the other spin-1/2 AFM trimer chain compound $Cu_3(P_2O_6OH)_2$ with no magnetic ordering [15, 16]. The exchange interactions ($J_2/J_1 \leq 0.3$), estimated from bulk susceptibility study, reveal that the compound can be considered as a good trimer system. Among the isostructural compounds from the series $AM_3P_4O_{14}$ (where $A$ = Ca, Sr, Ba, and Pb and $M$ = Ni, Co and Mn)[17-26], only the $AMn_3P_4O_{14}$ ($A$ = Sr and Ba) compounds with $S$=5/2 show 1/3 magnetization plateau. However, no magnetization plateau was found for isostructural compounds based on Co- ($S$ = 3/2) and Ni- ($S$ = 1) having low spin values [17, 22] which were predicted to be AFM trimer spin chain compounds [8]. However, later microscopic investigation of ground state properties by neutron powder diffraction suggested FM trimers in these compounds [17, 22]. For the compounds $ACo_3P_4O_{14}$ and $ANi_3P_4O_{14}$, magnetic long-range ordering temperatures were found to be higher (6.5 and 16 K, respectively [17, 22]) in contrast to the $T_N$ =2.2 K for $SrMn_3P_4O_{14}$ [25]. The exchange interactions in $SrMn_3P_4O_{14}$ were estimated by inelastic neutron scattering (INS) study [26] which revealed $SrMn_3P_4O_{14}$ as a good AFM trimer system with negligible intertrimer exchange interaction $J_2$. The appearance of the 1/3 magnetization plateau was reported due to quantum mechanical discrete energy levels of the magnetic eigenstates of the spin-5/2 AFM trimer [26]. On the other hand, no such study to estimate the exchange interactions is reported for the Co and Ni based compounds (where no 1/3 magnetization plateau exists), which could reveal microscopically the reason for the absence of 1/3 magnetization plateau. It is also not known how good or bad $ACo_3P_4O_{14}$ and $ANi_3P_4O_{14}$ compounds are as trimer systems which is defined by the $J_2/J_1$ ratio. For this purpose, inelastic neutron scattering is one of the best techniques to investigate the nature and strength of the exchange interactions; which can experimentally verify the theoretically predicted criteria for magnetization plateaus, as well as can define the degree of trimerization (how good or bad trimer system) from the $J_2/J_1$ ratio. The INS measures low-lying magnetic excitations and allows to determine directly the sign and strength of the possible exchange interactions from the energy/dispersion of the excitations.

Here, we report the results of INS study on the spin-1 trimer chain compound $CaNi_3P_4O_{14}$. The compound $CaNi_3P_4O_{14}$, belonging to the general formula $AM_3P_4O_{14}$ (where $A$ = Ca, Sr, Ba, and Pb and $M$ = Co, Mn, and Ni) [17-26], was initially predicted to show a unique long-range ferrimagnetic ordering due to the periodicity of the exchange interactions [11,26]. The trimer spin-chains in $CaNi_3P_4O_{14}$ were estimated, from the bulk magnetization data and Monte Carlo simulations, to be constructed of AFM $J_1$ and FM $J_2$ [8]. However, our recent neutron diffraction study [22] evaluated the nature of magnetic ground state as an uncompensated antiferromagnet where spins within trimer unit as well as within the chains have parallel arrangement suggesting the ferromagnetic type $J_1$ and $J_2$ in agreement with the results of recent first-principles calculation [23]. Our neutron diffraction results also reveal a 1D type magnetic behaviour above the $T_C$. Below the $T_C$ = 16 K, in spite of the $S$ = 1 trimer chain system, the compound $CaNi_3P_4O_{14}$ is found to be a good realization of a three-dimensional magnet with full ordered moment values of ~2 $\mu_B$/$Ni^{2+}$ at 1.5 K. Nuclear magnetic resonance (NMR) study and first principle calculations on $CaNi_3P_4O_{14}$ suggest an appearance of energy gap in the magnon excitation spectrum in the magnetically ordered state along with a finite orbital degree of freedom associated with the magnetic ordering [23]. NMR data also reveal the presence of anisotropy in the ordered state [23]. In the present work, we have employed inelastic neutron scattering to investigate the magnetic excitations and subsequently the spin Hamiltonian of $CaNi_3P_4O_{14}$. The present work aims to understand the reason for the absence of 1/3 magnetization plateau. The experimental results are corroborated by spin wave simulations, based on linear spin wave theory, which reveal the dominance of the both ferromagnetic intratrimer and intertrimer interactions within the given chains. The present study also reveals an antiferromagnetic coupling of the spin-chains and presence of a uniaxial single-ion-anisotropy. Our study brings out the role of interchain interactions and magnetocrystalline anisotropy on stabilization of the observed 3D long-range magnetic ordering in the trimer spin-chain compound $CaNi_3P_4O_{14}$. The present study provides an experimental verification of the theoretical predictions for the conditions to observe 1/3 magnetization plateau in spin-trimer chains. The present study also discusses the significance of the gapless magnetic excitations from the strongly correlated low-dimensional spin chains above the $T_C$.

**EXPERIMENTAL DETAILS:**

Polycrystalline samples of $CaNi_3P_4O_{14}$ (~20 g) were synthesized using a solid state reaction method in air. Stoichiometric mixture of NiO (99.99 %), $CaCO_3$ (99.99 %) and $(NH_4)_2HPO_4$ (99.99 %) was heated at 1000 ºC for 150 hrs. with several intermediate grindings. The phase purity and quality of the sample were verified by x-ray diffraction using a Cu $K_{\alpha1}$ radiation. Rietveld refinement of the experimental diffraction pattern, using the FULLPROF program [27], confirmed that the samples were single phase with monoclinic crystal structure [space group: $P2_1/c$; lattice parameters $a$ = 7.3301(2) Å, $b$ = 7.5798(2) Å, $c$ = 9.3929(11) Å, and $\beta$ = 111.989(1)°] which is consistent with literature reports [20, 22].

A low temperature diffraction pattern at 6 K was recorded by using the neutron powder diffractometer II ($\lambda$ = 1.2443 Å) at the Dhruva reactor, Trombay, India. The inelastic neutron-scattering (INS) measurements were performed on the high flux neutron time-of-flight instrument MERLIN at the ISIS facility of the Rutherford Appleton Laboratory, Didcot, U.K. The Merlin spectrometer has larger detector coverage in both the horizontal (~180º) and vertical (± 30º) scattering planes allowing measurements over large $Q$-regions of $S(Q,\omega)$ space. About 20-g powder sample was used for the INS measurements. The powder sample was placed in an envelope of thin aluminium foil (40 mm height and 140 mm length), which was rolled into cylindrical form and inserted inside a thin aluminium cylindrical can (diameter of 40 mm with wall thickness of 0.1 mm). The Al can was then mounted into a closed-cycle refrigerator having He-4 exchange gas for low temperature measurements. The straight Gd slit package was used in the Fermi chopper, which was phased (at a rotation speed of 350 Hz) to allow the recording of spectra with incident energies of 108, 40, 21, 13 meV, simultaneously, via the rep-rate multiplication method [28, 29]. The data were collected at several temperatures between $T$=4 and 100 K. A typical measurement time for each spectrum was ~3 hrs. Additional low energy spectra were collected with an incident energy of 8 meV (at a rotation speed of 300 Hz) at 4, 12 and 18 K with measuring time ~ 5 hrs. for each run. The Data were reduced using the MantidPlot software package [30]. The raw data were corrected for detector efficiency and time independent background following standard procedures. The spin wave simulations were carried out using the Spin-W program [31].

**RESULTS AND DISCUSSION:**

*Crystal structure -*

The low temperature crystal structure of $CaNi_3P_4O_{14}$ has been studied by powder neutron diffraction at 6 K. The derived structural parameters were used for the analysis of the low-temperature magnetic excitation spectra. The Rietveld refined diffraction pattern is shown in Fig. 1(a). The Rietveld analysis confirms that $CaNi_3P_4O_{14}$ crystallizes in the monoclinic symmetry (space group $P2_1/c$) with lattice parameters $a$ = 7.3006(8) Å, $b$ = 7.5462(9) Å, $c$ = 9.3465(10) Å, and $\beta$ = 111.96 (7)° at 6 K. The crystal structure is found to be similar to that at room temperature structure [22]. During the refinement, only nuclear phases of $CaNi_3P_4O_{14}$ and $Ni_3P_2O_8$ (a minor secondary phase; weight % ≈ 5%) have been considered. To avoid the influence of the magnetic signal, we have excluded the low $Q$ region ($< 2.1$ Å$^{-1}$) data from the refinement. All the atomic sites are considered to be fully occupied and kept fixed during the refinement. In the present crystal structure, there are two independent crystallographic sites for both Ni and P ions, seven independent crystallographic sites for oxygen ions, and a single crystallographic site for Ca ions.

The spin-chains in CaNi$_3$P$_4$O$_{14}$ are formed by edge-shared NiO$_6$ (Ni$^{2+}$; $3d^8$, $S = 1$) octahedra along the crystallographic $b$ axis [Fig. 1(b)]. Within a given chain, the special periodicity of the magnetic Ni$^{2+}$ ions Ni(1)-Ni(2)-Ni(1) at two crystallographic independent sites 4$e$ [Ni(1)] and 2$a$ [Ni(2)] is resulted into a spin-trimer structure ($J_1$-$J_1$-$J_2$) having intrachain exchange interactions $J_1$ [(between Ni(1) and Ni(2); $d_{Ni(1)-Ni(2)} = 3.096(6)$ Å] and $J_2$ [between Ni(1) and Ni(1); $d_{Ni(1)-Ni(1)} = 3.165(8)$ Å)] with different strengths. The superexchange pathways are Ni(1)-O2-Ni(2) and Ni(1)-O3-Ni(1), respectively. Such spin chains are separated by PO$_4$ tetrahedra suggesting weaker interchain exchange interactions. The possible interchain interaction, between two Ni ions from two adjacent chains having shortest distance $d_{Ni(1)-Ni(2)} = 4.663(5)$ Å, is through the superexchange pathway Ni(1)-O3-P1-O2-Ni(2)/ Ni(1)-O5-P2-O6-Ni(2) in the $bc$ plane. Moreover, other interchain interactions may also be present between the Ni ions, via PO$_4$ tetrahedra, having slightly longer distances $d_{Ni(1)-Ni(1)} = 5.036(6)$ Å, $d_{Ni(1)-Ni(2)} = 5.191(4)$ Å and $d_{Ni(2)-Ni(2)} = 6.0063(5)$ Å within the $bc$ plane and $d_{Ni(1)-Ni(1)} = 5.176(8)$ Å in the $ab$ plane, respectively.

*Spin wave excitations* -

The color-coded inelastic neutron scattering intensity maps of CaNi$_3$P$_4$O$_{14}$, measured on MERLIN at $T = 4$ K with various incident neutron energies $E_i = 8$, 13, 21 and 40 meV, are shown in Figs. 2(a)–2(d), respectively. All the observable magnetic scatterings are situated below ~ 6.5 meV. There are three bands of magnetic scatterings over ~1-2.5, ~2.8-4.6, and ~5.3-6.5 meV, respectively. The magnetic character of the scattering is evident from the decreasing intensity with increasing |$Q$|. The magnetic scatterings are found to be extended up to |$Q$| ~ 5 Å$^{-1}$. For the $E_i = 40$ meV spectrum, high energy scatterings (centred around 20 meV) that appear at high momentum transfer region (|$Q$| > 5 Å$^{-1}$) are due to the scattering from phonon excitations. Due to the polycrystalline nature of the samples, the scattering cross-section $S(|Q|, \omega)$ is powder average of the spin-spin correlation function $S(Q, \omega)$, and it does not carry the information regarding the direction of $Q$, however, preserves singularities arising in the density of states as a function of $E = \hbar\omega$. It is found that the powder neutron spectrum contains distinctive fingerprints of the Hamiltonian which can be readily compared to theoretical calculations to obtain approximate parameters.

We have also investigated the temperature variation of the magnetic excitation spectra over a wide temperature range between 4 and 100 K [Figs. 3 and 4]. Looking at the temperature evolution of the spectra, the flat magnetic scattering bands appear to be strongly affected by increasing temperature. The intensity of the low energy bands weakens quickly with increasing temperature. At the same time, the modes soften, and excitation spectra become gapless above the $T_C$. This implies that the low-|Q| scatterings are related to the spin wave excitations in the magnetic ordered state (long-range) of the material. Weak gapless magnetic excitations persist above the $T_C$ up to ~ 100 K [Figs. 3 and 4], which indicates the presence of short-range spin-spin correlations within the chains. Short-range spin-spin correlations were indeed found in neutron diffraction study as reported by us [22]. This may be noted that there are no phonon modes whose intensity increases with increasing temperature around the spin wave spectra. This makes our data clean and easy to analyse/explain without subtracting the phonon background.

In order to model the experimentally observed magnetic spectrum, we have calculated the spin wave dispersions, the spin-spin correlation function, and the neutron scattering cross section using the SpinW program [31]. The studied compound $CaNi_3P_4O_{14}$ contains only the magnetic ions $Ni^{2+}$ ($3d^8$, $S = 1$), and therefore, only interactions between the $Ni^{2+}$ ions need to be considered. Considering the trimer spin chain crystal structure, we have constructed the magnetic Hamiltonian with three different exchange couplings [shown in Fig. 1(b)] as

$$H = \sum_i J_1(\vec{S}_{3i-2}\cdot\vec{S}_{3i-1}) + J_1(\vec{S}_{3i-1}\cdot\vec{S}_{3i}) + J_2(\vec{S}_{3i}\cdot\vec{S}_{3i+1}) + J_3 \sum_{ij}(\vec{S}_i\cdot\vec{S}_j) + \sum_i D(S_i^z)^2 \quad (1)$$

where $J_1$ and $J_2$ are the nearest neighbour intrachain exchange interaction between Ni(1) and Ni(2) ions and between Ni(1) and Ni(1) ions, respectively, within a chain. The $J_3$ is the interchain interactions between two Ni ions from two adjacent chains. Out of many possible interchain interactions having direct distances between 4.663 and 6.0063 (having similar superexchnage pathways via Ni-O-P-O-Ni; as discussed in the crystal structure section), we have considered the one which is having shortest direct distance $d_{Ni(1)-Ni(2)} = 4.663(5)$ Å [through the superexchange pathways Ni(1)-O3-P1-O2-Ni(2)/Ni(1)-O5-P2-O6-Ni(2) in the *bc* plane] [Fig. 1(b)]. The fourth term in Eq. (1) is due to single-ion-anisotropy parameter *D* which originates from the crystal field of the surrounding oxygen ions in a $NiO_6$ octahedral environment. The anisotropy parameter *D* induces a small gap between ground state and excited states, as found in the experimentally measured spectrum [Fig. 5(c)]. For simplicity, we have considered the same *D* value for both Ni(1) and Ni(2) sites. As the Ni-3d wave functions are very localized, long-range Ni-Ni interactions are neglect. For the simulation of the spin wave spectra we used the canted magnetic structure as reported earlier by us [22], having a dominated spin component along the *c* axis and a weak spin-component along the *b* axis. By adjusting the values of the exchange interactions ($J_1$, $J_2$, and $J_3$) and the anisotropy parameter *D*, a good solution could be found (Table I), and the corresponding simulated excitation pattern is depicted in Fig. 5(b). The simulated energy and momentum cuts are shown in Figs.

5(f) and (g) along with that obtained from the measured pattern. An excellent agreement is obtained between the experimentally measured and the model calculated patterns. The major features of the excitation spectra are as follows: (i) significant bandwidth of the three excitation bands with intermediate gaps [Fig. 5(f)], (ii) a dispersive nature of the bottom edge of the lowest energy band [Fig. 5(d-e)], (iii) a modulation of the intensity of the lower energy band as a function of momentum transfer $|Q|$ [Fig. 5(g)], and (iv) the presence of an energy gap of about 1 meV between the ground state and the lowest edge of the excitation spectrum as confirmed by the constant energy cuts [Fig. 5(c)].

Our spin wave simulations reveal that the observed three energy-bands correspond to the three spin wave dispersion modes with distinguished energy ranges. The widths of the energy bands signify the relative strengths of the intrachain interactions ($J_1$ and $J_2$). For a model with equal intrachain exchange interactions i.e., $J_1 = J_2$, the excitation spectra is continuous in energy i.e., a single band formation from the lowest energy to highest energy (over 1 - 6.5 meV). In this case, there is no gap present between the three dispersion modes. Any difference in the strength of $J_1$ and $J_2$ creates energy gaps between the dispersion modes resulting into observed three discrete energy bands. The widths of the energy bands and the value of the gaps are proportional to the ratio between the $J_1$ and $J_2$. The spin wave simulations also reveal that the antiferromagnetic interchain interaction $J_3$ which lies in the bc plane weakly modifies the gaps between the bands. The $J_3$ also results into a dispersion along the c axis. The main contribution from the $J_3$ appears at the bottom of the lowest energy band where a clear dispersion is visible in the experimentally measured spectrum at low $Q$ region. The fitting suggests that both intratrimer ($J_1$) and intertrimer ($J_2$) interactions are ferromagnetic and have almost similar strength; $J_1$ = -1.3±0.05 meV and $J_2$ = -1.05±0.05 meV, respectively. The strength of the interchain interaction $J_3$ = 0.9±0.1meV is found to be weaker and an antiferromagnetic type. The spin gap of ~ 1 meV confirms the presence of a single ion anisotropy. The anisotropy is found to be along the c axis i.e., along the moment direction (the easy axis). The fitted value of the single ion anisotropy parameter $D$ is found to be -0.25 meV. In summary, the fitting of the coupled trimer spin-chain model parameters to the experimental data reveals an essential information regarding the magnetism of $CaNi_3P_4O_{14}$. The derived values of the exchange constants show that $CaNi_3P_4O_{14}$ is composed of ferromagnetic trimers that are coupled ferromagnetically along the chain direction (b axis); while such chains are coupled antiferromagnetically.

A comparison between the derived values of exchange coupling parameters from our INS data with the values reported from the DFT calculations is given in Table-I. The signs of all the three exchange interactions obtained from the experiment and DFT calculations are found to be consistent. The dominating interactions are found to be ferromagnetic within the chains through the superexchange pathways Ni(1)-O2-Ni(2) /Ni(1)-O6-Ni(2) and Ni(1)-O3-Ni(2). The ferromagnetic intratrimer and intertrimer interactions in $CaNi_3P_4O_{14}$ could be understood by the Goodenough-Kanamori rule [32, 33]. For both the cases, superexchange interactions are between two half-filled Ni-$e_g$ orbitals via superexchange pathways Ni-O-Ni having bond angles (∠ Ni-O-Ni ~99º/94º for $J_1$ and ∠ Ni-O-Ni ~100º for $J_2$) close to 90º. According to the Goodenough-Kanamori rule, the superexchange interactions between two half-filled orbitals connected by a superexchange angle close to 90º always favor a weak ferromagnetic interaction due to the possibility of a direct overlap between two half-filled orbitals (here Ni-$e_g$ orbitals). Although the signs of the exchange interactions, obtained from both experimental measurements and the DFT calculations, are consistent, their values are found to be significantly different. Experimental data reveal the ratio $J_2/J_1$ ~ 0.81 as compared to ~ 0.38-0.52 obtained from the DFT calculations. Moreover, significant discrepancy has been found for the interchain interaction $J_3$. The $J_3$ was predicted, by DFT calculations (by both the GGA and GGA+$U_{eff}$), to be the strongest interaction ($J_3/J_1$ ~1.8-1.9) in spite of the longer superexchange pathways via Ni-O-P1-O-Ni. However, experimental data reveal that the long-range $J_3$ interaction has a weaker strength ($J_3/J_1$ ~ 0.7). The spin wave spectra for the values predicted by the DFT calculations (both the GGA and GGA+$U_{eff}$) are simulated and shown in Fig. 6. Although the simulated powder averaged spectra reveal three excitation bands, their individual as well as overall energy range, band width and intensities are significantly different from the experimental spectra for $CaNi_3P_4O_{14}$ [Fig. 2]. This demands further careful DFT-based first-principles calculations, especially proper choice of Hubbard on-site Coulomb correlations and the Hunds exchange parameter ($J_H$), and proper estimation of charge transfer energies between different orbitals, with a more accurate crystal structure, and higher plane-wave cut-off energy.

The presence of the energy gap ~ 1 meV in the magnon excitation spectrum below $T_C$= 16 K is consistent with the reported NMR data that showed a thermally activation behaviour of the spin-lattice relaxation rate $1/T_1$ for the P sites [23]. Our spin wave calculations reveal a single-ion-anisotropy with the anisotropy parameter $D$ = -0.25(2) meV. The anisotropy axis is found to be along the $c$ axis which is the easy axis. The GGA+U+SOC based DFT calculations [23] reveal that the magnetocrystalline anisotropy is weak and favours easy-axis single-ion magnetocrystalline anisotropy, which are consistent with our experimental data. The INS data reveal that the energy gap disappears above the $T_C$ [Fig. 4(c)]. A significant broadening of the excitation bands is also evident above the $T_C$ [Fig. 4(b)]. The sharp change in the single-ion anisotropy indicates that the stability of the 3D magnetic ordering in $CaNi_3P_4O_{14}$ is ascribed to the local magnetic anisotropy in addition to the interchain interactions $J_3$.

The presence of weak gapless magnetic excitations that persist up to ~ 100 K, is consistent with the previous neutron diffraction study [22] which revealed 1D short-range spin-spin correlations at $T > T_C$. The integrated intensity for the highest energy band is shown in Fig 4(d) as a function of temperature. A significant intensity has been found to persist above the $T_C$ = 16 K which is attributed to the magnetic excitation from the strongly correlated spin-chains. Our linear spin wave model indeed reveals a strong coupling along the chains with $J_3/J_1 \sim 0.69$. The main broad diffuse peak in the neutron diffraction patterns appears over $Q$ = 0.4-1.2 Å$^{-1}$ and centred around $Q$ = 0.7 Å$^{-1}$. In the present INS data, gapless magnetic excitations appear over a similar $Q$ region above the $T_C$, revealing the origin as the 1D short-range magnetic ordering and spin fluctuations in the 1D magnetic ordering. The integrated intensity of the diffuse magnetic neutron scattering data also shows a maximum in the vicinity of the $T_C$. The NMR data [23] also revealed a change of the ratio $(1/T_1)/(1/T_2)$ ($T_2$ = spin-spin relaxation rate) (which actually gives the ratio of dynamical susceptibility $\chi_\perp$ and $\chi_\parallel$) below 50 K. Such one dimensional spin-spin correlations [34] may lead to exotic physical properties, such as magneto-dielectric coupling above 3D ordering temperature as reported for the spin-chain compound $Ca_3Co_2O_6$ [35], and hence opens up opportunities for further studies.

Now we compare the exchange parameters estimated for $CaNi_3P_4O_{14}$ with those of the isostructural compound $SrMn_3P_4O_{14}$ [24-26] which show 1/3 magnetization plateau. For $SrMn_3P_4O_{14}$, discrete magnetic excitations were reported with peaks at 0.46, 0.68, and 1.02 meV in contrast to the spin wave excitation for the present compound $CaNi_3P_4O_{14}$. The dispersion modes for $CaNi_3P_4O_{14}$ along the different crystallographic directions are shown in Figs. 7 (a). For $SrMn_3P_4O_{14}$, the excitations were assigned to the transitions between discrete energy levels [26]. The intratrimer exchange interaction $J_1$ was estimated to be antiferromagnetic with a strength of 0.29 meV, in contrast to the ferromagnetic $J_1$ with a value of -1.3 meV in the present compound $CaNi_3P_4O_{14}$. For $SrMn_3P_4O_{14}$, the intertrimer exchange interaction $J_2$ was concluded to be antiferromagnetic and very weak. Therefore, $SrMn_3P_4O_{14}$ could be considered as a good spin trimer compound. The 1/3 magnetization plateau in $SrMn_3P_4O_{14}$ originates due to the discrete energy levels of the magnetic eigenstates of the spin-5/2 AFM trimer [24, 26] where the energy gap between the plateau state (total spin 5/2) and the higher state (total spin 7/2) prevents the magnetization to be increased with magnetic field. The long-range magnetic ordering in $SrMn_3P_4O_{14}$ was explained by the competition between nearest neighbour and next nearest neighbour exchange interactions within a given chain [36] and the discrete energy levels persist even in the 3D ordered state. On the other hand, the topology of the exchange interactions in the present compound $CaNi_3P_4O_{14}$ reveals ferromagnetic trimer chains (FM-FM-FM) with strong interchain coupling. The spin configuration of $CaNi_3P_4O_{14}$ leads to a ferromagnetic chain, whose excitations are dispersive spin waves as found in our experimental spectra [Figs. 5 and 7]. In this case no magnetization plateau is expected. As outlined in the introduction, for $S$ = 1 FM trimer chains, the 1/3 and 2/3 magnetization plateaus are expected only when the FM trimers ($J_1$ FM) are coupled antiferromagnetically ($J_2$ AFM) i.e., spin chain with FM-FM-AFM periodicity, and the ratio $J_2/J_1$ is above a critical value [11]. The required conditions are not fulfilled by the present $CaNi_3P_4O_{14}$ compound. A variation of the dispersion curves with the $J_2/J_1$ is shown in Fig. 7(c) which reveals that with decreasing $J_2$ value the band widths of the dispersion modes decrease and subsequently the gaps between them increase. For $J_2 = 0$, the modes become dispersion less and as a result the excitations consist of discrete energy states which is one of the conditions to have the 1/3 magnetization plateau as found for the isostructural $SrMn_3P_4O_{14}$ compound [26]. Therefore, with a proper crystal modification, the magnetization plateaus state may be possible to introduce in a spin trimer chain compound like $CaNi_3P_4O_{14}$. Moreover, the present experimental study reveals a strong interchain exchange interaction ($J_3/J_1 \sim 0.7$) in $CaNi_3P_4O_{14}$ which leads to the higher 3D long-range ordering temperature $T_C = 16$ K as compared to that for the isostructural compounds ($T_N = 6.5$ and 2.2 K for Co and Mn ($J_3/J_1 < 10^{-3}$) based compounds, respectively). Figure 7(b) shows the effect of interchain interaction $J_3$ on the excitation spectra. The strong interchain exchange interaction $J_3$ also results into the full ordered moment values of 1.98 and 1.96 $\mu_B/Ni^{2+}$ at 1.5 K [theoretical value is 2 $\mu_B/Ni^{2+}$ ($S = 1$)] for the present compound. The spin-1 chain system, $CaNi_3P_4O_{14}$ is therefore found to be a good realization of a 3D magnetic system below $T_C = 16$ K with a negligible effect of quantum fluctuations.

**SUMMARY AND CONCLUSION**

We have performed inelastic neutron scattering measurements on the spin-1 trimer chain compound $CaNi_3P_4O_{14}$, which reveal gapped dispersive spin wave excitations at low temperatures ($T < T_C$). The magnetic excitations partially survive at temperatures (up to ~100 K) much above the $T_C = 16$ K (of long range 3D order) indicating the dominating 1D magnetic interactions. We performed the quantitative analysis of the observed magnetic excitation spectra using the linear spin-wave calculations. The fitting of the coupled trimer spin-chain model parameters to the experimental data reveals essential information regarding the magnetism of $CaNi_3P_4O_{14}$. Our spin wave analysis indicates that the stronger exchange interactions are within the chain along the $b$ axis. The signs of the exchange interactions are in agreement with the reported DFT calculations. However, disagreements have been found on the strength of the interactions. The strengths of the both ferromagnetic intratrimer ($J_1$) and intertrimer ($J_2$) nearest neighbour exchange interactions are found to be similar ($J_2/J_1 \sim 0.81$) which is in contrast to the predicted values ($J_2/J_1 \sim 0.38$-0.52) by the DFT calculations. Moreover, the DFT study predicted a strongest interchain interaction ($J_3/J_1 \sim 1.8$-1.9) which is in contrast to our inelastic neutron scattering results that reveal a weaker $J_3$ ($J_3/J_1 \sim 0.69$). The derived weaker $J_3$ is consistent with the spin chain type crystal structure. Nevertheless, the relatively strong value of the $J_3$ leads to the conventional 3D type magnetic ordering behaviour below the $T_C$ with full ordered moment for $Ni^{2+}$ ions at 1.5 K in the present spin-1 trimer chain compound $CaNi_3P_4O_{14}$. Furthermore, the observation of a spin gap below the $T_C$ is due to a weak single-ion-anisotropy ($D = -0.25$ meV). The sharp change of single-ion anisotropy across the $T_C$ reveals that the stability of the 3D magnetic ordering in $CaNi_3P_4O_{14}$ is ascribed to the local magnetic anisotropy in addition to the interchain interactions. The present study determines the spin Hamiltonian for the trimer spin-chain compound $CaNi_3P_4O_{14}$; that not only explains the observed physical properties in $CaNi_3P_4O_{14}$ but also explains why the 1/3 quantum magnetization state is present only for few members across the series of the isostructural compounds. The present study can foster research on the magnetic excitations in this class of spin-chain materials with multiple intrachain interactions and would generate theoretical interest in the development of a more realistic model to understand the complex magnetic behaviors. Besides, the observed gapless magnetic excitations up to a higher temperature of $T/T_C \sim 6$ generates interest to look for unexplored physical properties, such as magneto-dielectric and magneto-thermal properties.

**ACKNOWLEDGMENT:**

AKB and SMY thank K. Ghoshray for scientific discussion on the present compound. AKB also acknowledges the help received from H. C. Walker on the determination of the instrumental resolution of Merlin spectrometer.

**TABLE I**: Possible pathways for intratrimer, intertrimer, and interchain exchange interactions $J_1$, $J_2$ and $J_3$. The Ni-Ni direct distances, bond lengths, and bond angles for the exchange interactions in $CaNi_3P_4O_{14}$ at 6 K. The fitted values of the exchange interactions $J$'s and anisotropy parameter ($D$) from inelastic neutron scattering spectra at 4 K. All the values of exchange interactions are in meV.

| Exchange interaction | Ni…Ni direct distance (Å) and exchange pathways | Bong lengths (Å) | Bond angles (º) | Values (INS) (meV) | Values GGA (meV) [23] | Values GGA+$U_{eff}$ +(meV) [23] |
|---|---|---|---|---|---|---|
| $J_1$ | Ni(1)…Ni(2) = 3.096(6) Ni(1)-O2-Ni(2)/ Ni(1)-O6-Ni(2) | Ni(1)-O2 = 2.051(8) Ni(2)-O2 = 2.018(8) Ni(1)-O6 = 2.137(11) Ni(2)-O6 = 2.097(7) | Ni(1)-O2-Ni(2) = 99.1(3) Ni(1)-O6-Ni(2) = 94.0(3) | -1.30±0.05 [FM] | -2.85 | -1.54 |
| $J_2$ | Ni(1)…Ni(1) = 3.165(8) Ni(1)-O3-Ni(1) | Ni(1)-O3 = 2.051(9), 2.088(10) | Ni(1)-O3-Ni(1) = 99.8(3) | -1.05±0.05 [FM] | -1.49 | -0.578 |
| $J_3$ | Ni(1)…Ni(2) = 4.663(5) Ni(1)-O3-P1-O2-Ni(2)/ Ni(1)-O2-P1-O7-Ni(2) | Ni(1)-O3 =2.088(10) P1-O3 = 1.452(12) P1-O2 =1.515(13) Ni(2)-O2 =2.018(8) Ni(1)-O2 =2.051(8) P1-O2 =1.515(13) P2-O7 =1.552(14) Ni(2)-O7 =2.047(8) | | 0.9±0.1 [AFM] | 5.63 | 2.66 |
| $D$ | | | | -0.25±0.02 | | |

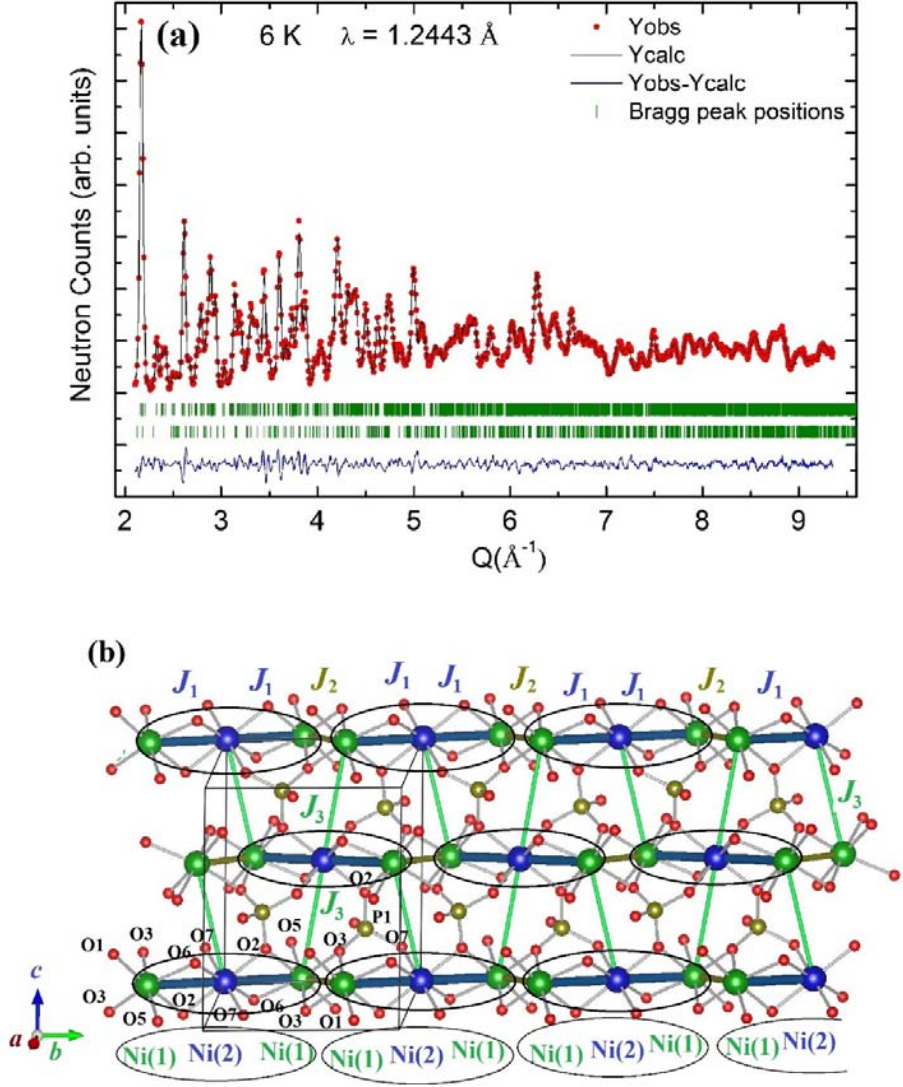

**Fig. 1:** (a) Experimentally observed (circles) and calculated (line through the data points) neutron diffraction patterns for $CaNi_3P_4O_{14}$ at 6 K over $Q$ = 2.1-9.4 Å$^{-1}$. The difference between observed and calculated patterns is shown by the solid blue line at the bottom. The vertical bars indicate the positions of allowed nuclear Bragg peaks for the main phase $CaNi_3P_4O_{14}$ (top row) and the minor secondary phase $Ni_3P_2O_8$ (bottom row), respectively. (b) A schematic view of the exchange interactions in $CaNi_3P_4O_{14}$. The Ni(1), Ni(2), P and O ions are shown by green, blue, yellow and red spheres, respectively. Ca ions are omitted for clarity. The intrachain exchange interactions between Ni(1) and Ni(2) ions are denoted by $J_1$, whereas, between Ni(1) and Ni(1) ions are denoted by $J_2$. The interchain interaction between two Ni ions having shortest distance ($d_{Ni(1)-Ni(2)}$ = 4.663 Å) is shown by $J_3$. The individual trimer units are marked by ellipsoid.

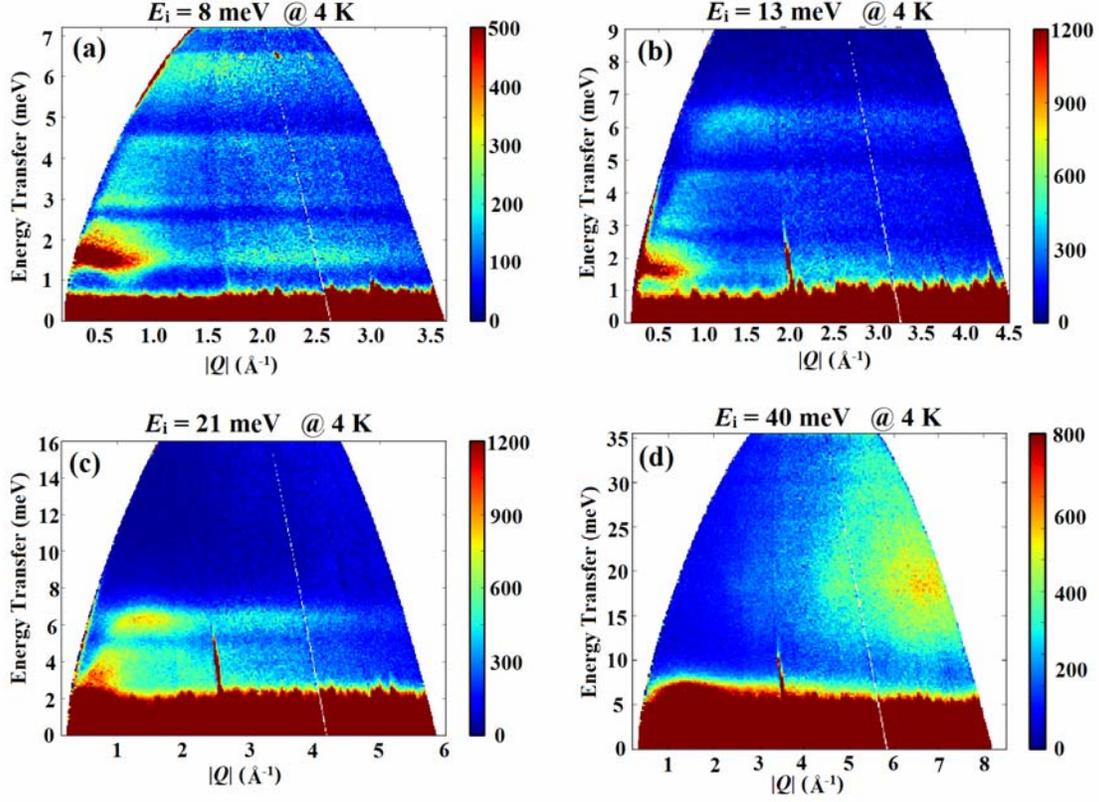

**FIG 2:** The 2D Color map of the INS intensity of $CaNi_3P_4O_{14}$ as a function of energy transfer ($\hbar\omega$) and momentum transfer ($|Q|$), measured at 4 K on the Merlin spectrometer, with incident neutron energy of $E_i$ = 8, 13, 21, 40 meV. The color scales show the scattering intensity $S(|Q|,\omega)$ in an arbitrary unit. The sharp features with strong intensities (on the lower edge of the spectra, over 5-7 meV for $E_i$ = 8 meV, over 1-4 meV for $E_i$ = 13 meV and over 2-8 meV for $E_i$ = 21 meV) are experimental artefacts caused by defective detector elements. The sharp feature that is observable at the lower energies around $|Q|$ ~ 2 Å-1 for $E_i$ = 13 meV, $|Q|$~ 2.5 Å-1 for $E_i$ = 21 meV and $|Q|$ ~ 3.5 Å-1 for $E_i$ = 40 meV could be due to some electronic instability in data accusation units (DAE) in time channels which automatically becomes stable with time [high temperature patterns above 25 K (Fig. 3)].

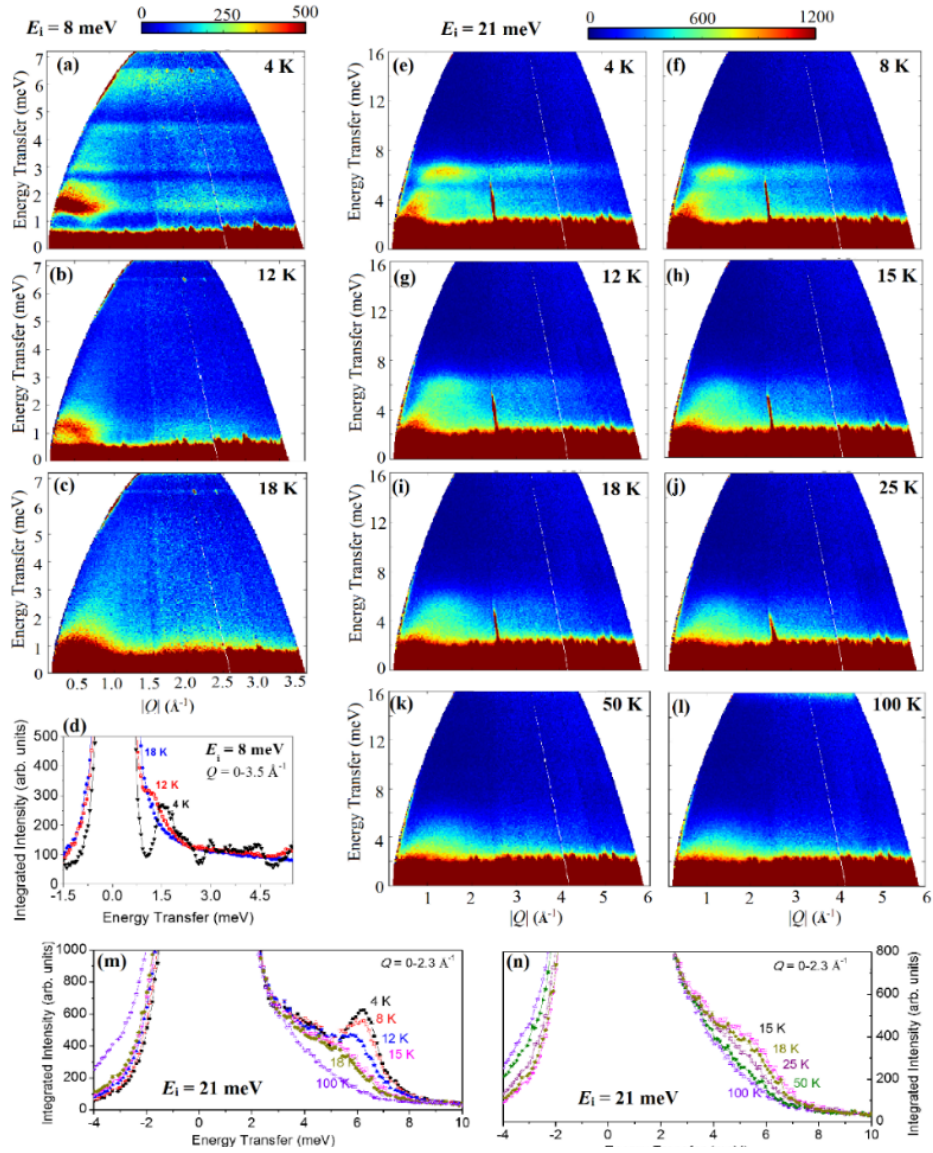

**FIG 3:** The 2D Color map of the INS intensity of $CaNi_3P_4O_{14}$ as a function of energy transfer ($\hbar\omega$) vs momentum transfer ($|Q|$) measured on the Merlin spectrometer. (a-c) The inelastic spectra measured with incident neutron energy of $E_i$ = 8 meV at 4, 12 and 18 K, respectively. (e-l) The INS spectra measured with incident neutron energy of $E_i$ = 21 meV at 4, 8, 12, 15, 18, 25, 50 and 100 K, respectively. The color scales show the scattering intensity $S(|Q|,\omega)$ in an arbitrary units. (d, m and n) The intensity vs energy transfer curves for $E_i$ = 8 and 21 meV, respectively. For $E_i$ = 8 meV, the intensities were obtained by integrations over $|Q|$ = 0-3.5 Å$^{-1}$ and for $E_i$ = 21 meV, the integrations were performed over $|Q|$ = 0-2.3 Å$^{-1}$).

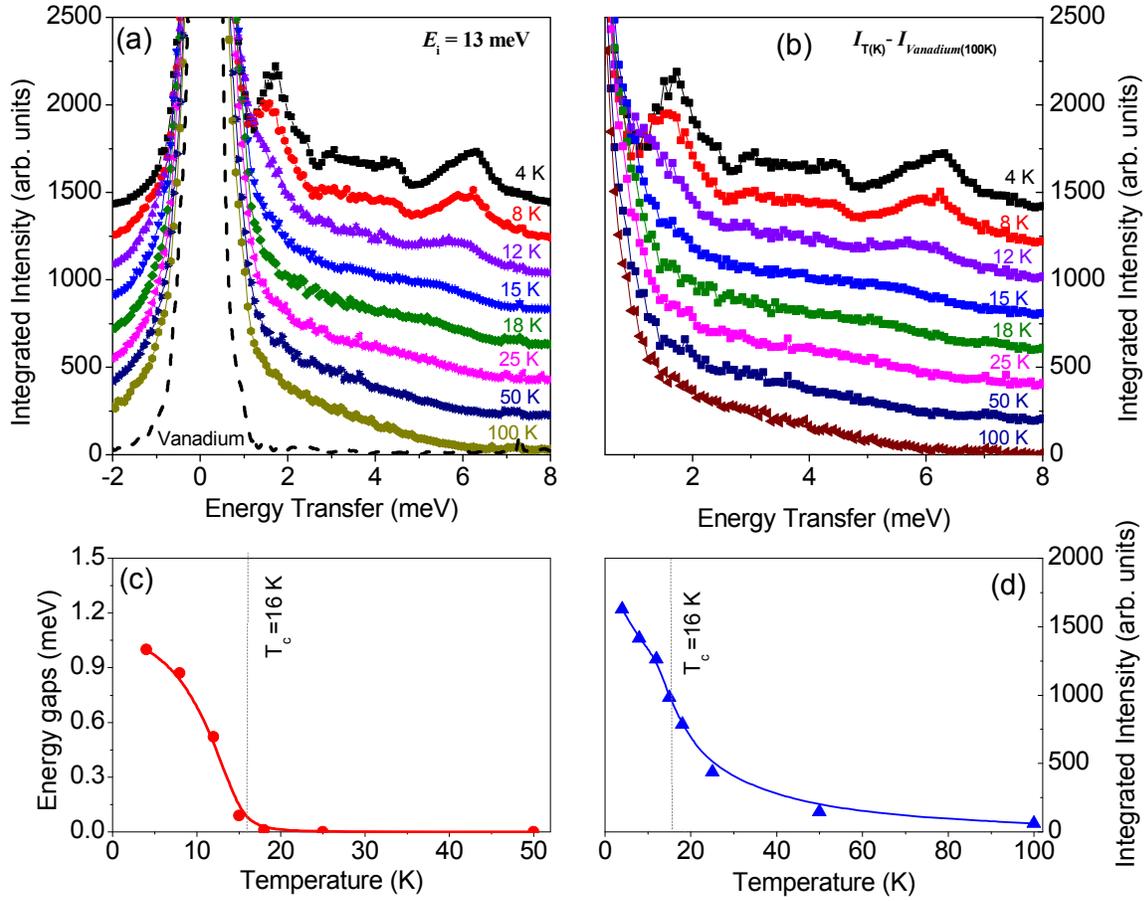

**FIG 4:** (a) The measured intensity vs energy transfer curves for $E_i$ = 13 meV at different temperatures. The instrumental profile is shown by the dashed line (vanadium curve). The intensities were obtained by integrations over $|Q|$ = 0-1.8 Å$^{-1}$. The curves are shifted vertically for clarity. (b) The measured curves after subtraction of V curve at 4, 8, 12, 15, 18, 25 and 50 K. (c) and (d) The temperature variation of the energy gaps and the integrated intensity, respectively.

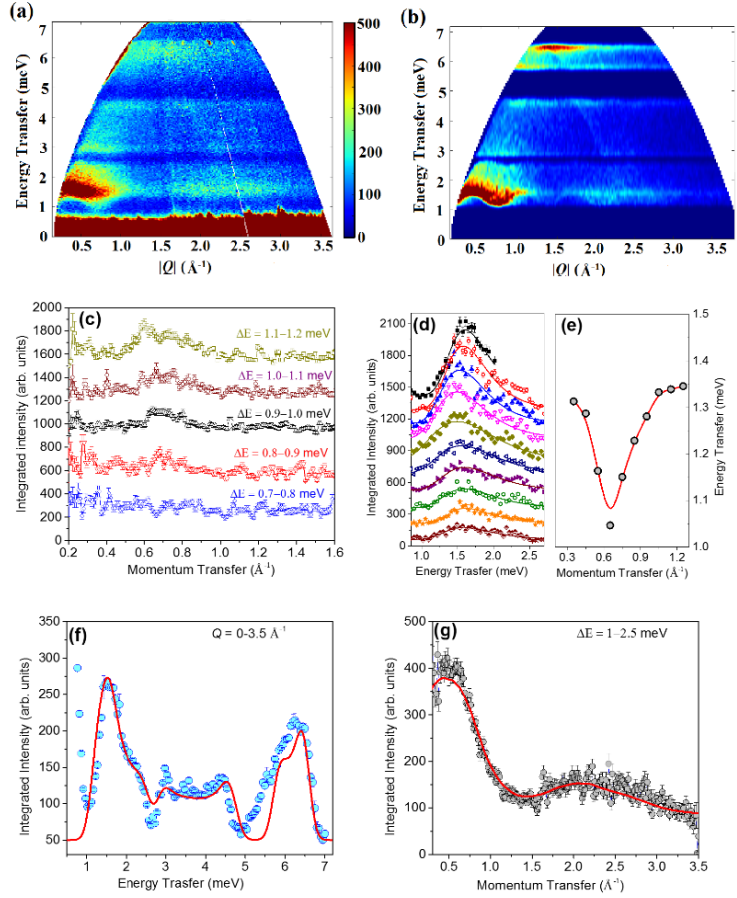

**FIG. 5:** The (a) experimentally measured (at 4 K) and (b) simulated (by the SpinW program) spin wave excitation spectra. The strong intensity over 5-7 meV on the lower edge of the experimental spectrum is an artefact caused by defective detector elements. (c) The constant energy cuts as a function of momentum transfer; summed over 0.7-0.8, 0.8-0.9, 0.9-1.0, 1.0-1.1 and 1.1-1.2 meV, respectively. (d) The constant momentum transfer |Q| cuts over the lower edge of the lowest energy excitation band illustrating the dispersion of the excitation mode. The solid lines are the guide to the eyes. Cuts are taken for |Q|= 0.35, 0.45, 0.55, 0.65, 0.75, 0.85, 0.95, 1.05, 1.15, and 1.25 Å$^{-1}$ with a width of 0.1 Å$^{-1}$ (|Q|±0.05 Å$^{-1}$). (e) The momentum dependence of the lower edge of the lowest energy band. The experimental scattering intensity as a function of (f) energy transfer (integrated over |Q| range 0–3.5 Å$^{-1}$) and (g) momentum transfer (integrated over ΔE = 1-2.5 meV), respectively. The spin wave calculated intensities (red solid lines) are also plotted for comparisons. To match with the experimental intensity, a constant scale factor to the calculated intensity has been applied in addition to a constant background. The addition intensity for the experimental spectra over 5-7 meV in (f) appears from the defective detector elements (for details see caption of Fig. 2).

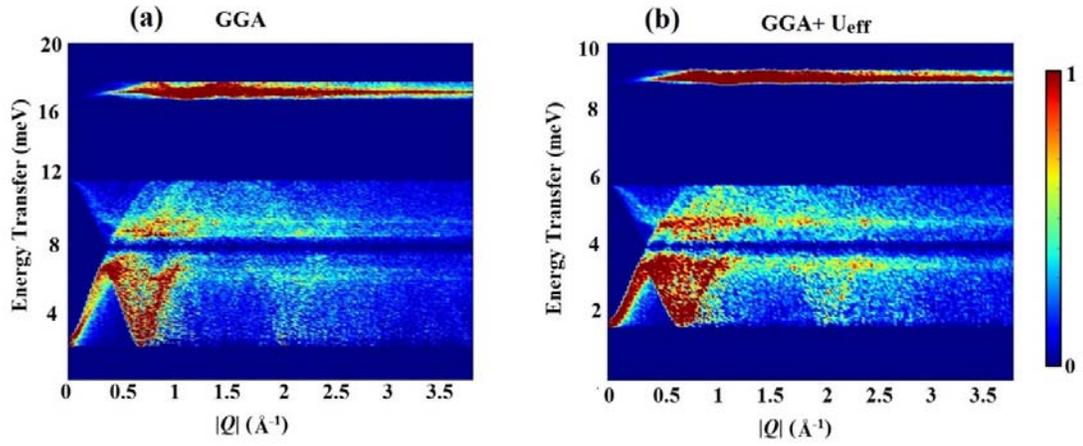

**FIG 6:** The powder averaged spin-wave excitation spectra for the values, predicted by the DFT calculation for $CaNi_3P_4O_{14}$, (a) GGA model $J_1 = -2.85$, $J_2 = -1.49$, $J_3 = 5.63$ and $D = -0.25$ meV, and (b) GGA+$U_{eff}$ model $J_1 = -1.54$ $J_2 = -0.578$, $J_3 = 2.66$ and $D = -0.25$ meV. Color bar represents the intensity in the arb. units.

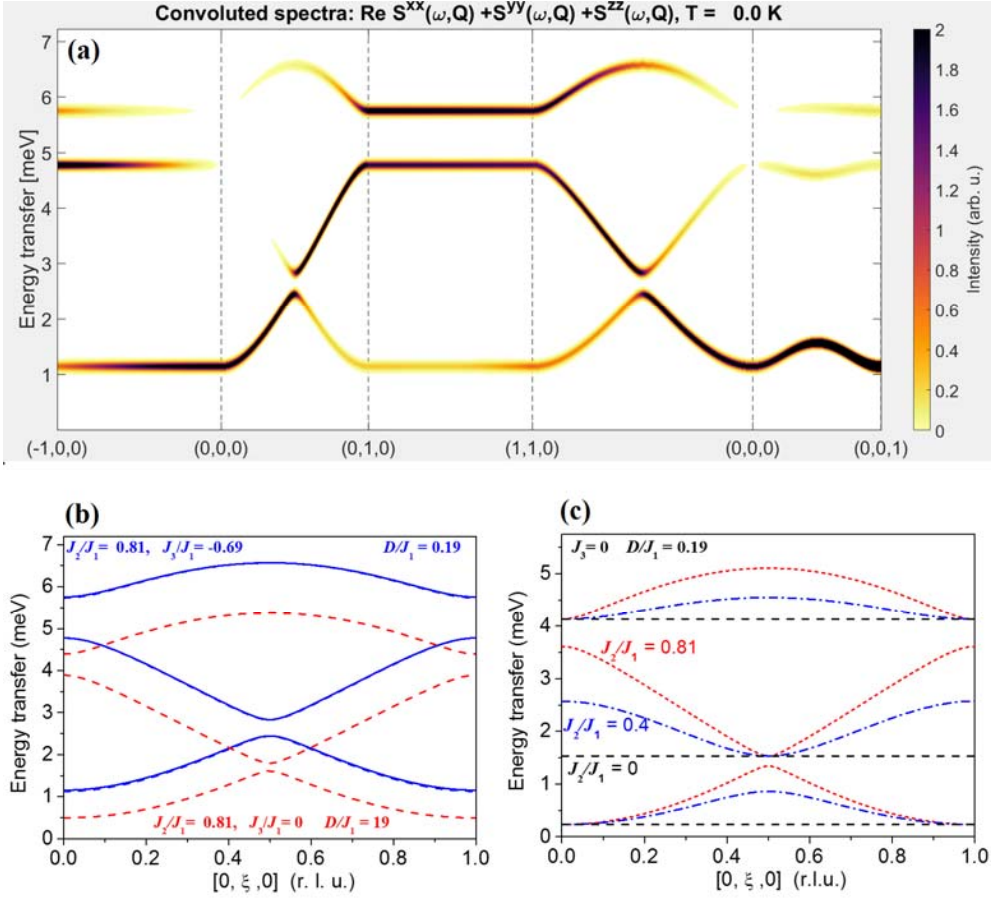

**FIG 7:** (a) The simulated dispersion curves along the different crystallographic directions with the derived parameters $J_1 = -1.3$, $J_2 = -1.05$, $J_3 = 0.9$ and $D = -0.25$ meV. The intensity variation of the dispersion patterns shown by the color map. (b) The dispersion curves (solid blue lines) along the chain direction (*b* axis) with the derived parameters $J_2/J_1 = 0.81$, $J_3/J_1 = -0.69$ and $D/J_1 = 0.19$ meV. The dashed curves represent the nature of the dispersion curves in the absence of the interchain interactions ($J_3 = 0$). (c) The variation of the dispersion curves with $J_2/J_1$ and $J_3 = 0$. Discrete energy levels are evident for $J_2 = 0$.